\documentclass[3p,times,procedia]{elsarticle}
\usepackage{ecrc}

\volume{00}

\firstpage{1}

\journalname{Physics Procedia}

\runauth{V.A. Gani et al.}

\jid{phpro}

\usepackage{amssymb}

\usepackage[figuresright]{rotating}

\begin{document}

\begin{frontmatter}



\dochead{Conference of Fundamental Research and Particle Physics, 18-20 February 2015, Moscow, Russian Federation}

\title{Two-dimensional manifold with point-like defects}


\author[a,b]{V.A. Gani\corref{cor1}}
\ead{vagani@mephi.ru}
\author[a]{A.E. Dmitriev}
\author[a]{S.G. Rubin}

\address[a]{National Research Nuclear University MEPhI (Moscow Engineering Physics Institute), Kashirskoe Shosse 31, Moscow, 115409, Russia}
\address[b]{National Research Center Kurchatov Institute, Institute for Theoretical and Experimental Physics, Bolshaya Cheremushkinskaya street 25, Moscow, 117218, Russia}

\begin{abstract}
We study a class of two-dimensional compact extra spaces isomorphic to the sphere $S^2$ in the framework of multidimensional gravitation. We show that there exists a family of stationary metrics that depend on the initial (boundary) conditions. All these geometries have a singular point. We also discuss the possibility for these deformed extra spaces to be considered as dark matter candidates.
\end{abstract}

\begin{keyword}
extra dimensions \sep multidimensional gravity \sep Planck mass \sep multiverse \sep dark matter

\PACS 04.50.-h \sep 95.35.+d

\end{keyword}
\cortext[cor1]{Corresponding author. Tel.: +7-916-630-4176.}
\end{frontmatter}




\vspace*{-8pt}
\section{Introduction}
\label{S:1}

Many problems of modern cosmology and of the Standard Model can be solved within multidimensional gravity. Strings, branes, multidimensional black holes and other objects have great potential to describe various physical phenomena. However, some questions can be solved by using the idea of extra dimensions only.

The standard starting point when one introduces extra dimensions is
to assume them to be in a maximally symmetric configuration.
This assumption makes it possible to obtain many interesting results \cite{2008IJMPD..17..785C,2011PhR...497...85M,2011PhRvD..84d4015B}.
On the other hand, there is no universal reason to believe the space-time geometry
to be really simple \cite{1991PhLB..259...38L,Lindebook}. The space-time
foam can be able to produce very different geometries, leading to
random initial conditions (at different points of the space-time foam) that
are a key point defining the properties of the extra dimensions.

This work considers different stationary geometries of
a compact two-dimensional extra space isomorphic to the $S^2$ sphere.
The Lagrangian of the model is a nonlinear function of the scalar curvature
that allows to stabilize the extra space size.
Particularly interesting is the setup where at each point of a bounded domain of the physical three-dimensional space, the corresponding two-dimensional extra space
has a deformed geometry. As we show, this space domain has an extra vacuum energy and could be considered as a dark matter particle. We have also estimated the vacuum energy as a function of the extra space deformation and discussed the scattering cross sections of such domains scattering off regular matter particles. Our estimate
appears to fulfill the experimental constraints.

\section{Two-dimensional extra space}
\label{S:2}

We will assume the extra space size to be small and its geometry having quickly stabilized after the Universe was born. We consider a manifold with the metric
\begin{equation}\label{metric}
ds^2 = g_{\mu\nu}(x)dx^{\mu}dx^{\nu} + G_{ab}(x,y)dy^a dy^b ,
\end{equation}
where $g_{\mu\nu}(x)$ and $G_{ab}(x,y)$ are the metric tensors of the four-dimensional space-time (the ``main space'') and of the extra space, respectively. In this paper we consider a metric uniform with respect to the coordinates of the main space, $G_{ab}(x,y)=G_{ab}(t,y)$. Moreover, the extra space metric stabilizes with time, i.e.\ $G_{ab}(t,y)\longrightarrow G_{ab}(y)$ at $t\rightarrow\infty$.

According to (\ref{metric}), the scalar curvature $R$ is the sum of the Ricci scalars of the main space $R_4$ and of the extra space $R_2$: $R=R_4+R_2$, $R_4\ll R_2$. Under this assumption we can write $R_4=\epsilon(x,y)R_2$, where $\epsilon(x,y)\ll 1$ for all $x$ and $y$, with the intention to perform later an expansion in powers of the small parameter.

We start from the action (see, e.g., \cite{nojiri01,sokolowski01})
\begin{equation}\label{act1}
S=\frac{m_D^{D-2}}{2}\int d^4 x d^2 y \sqrt{|G(y)g(x)|}f(R),\quad f(R)=\sum\limits_k {a_k R^k},
\end{equation}
where $a_1=1$ and $a_k$ are arbitrary constants when $k\neq 1$.
Upon expansion of $f(R)$ in powers of $\epsilon$, the action becomes
\begin{equation}
S\simeq \int d^4x  \sqrt{|g(x)|}\left[\frac{M^2 _{Pl}}{2}R_4 +  \frac{m_D ^{D-2}}{2}\int d^2 y \sqrt{|G(y)|} f(R_2)\right],
\end{equation}
where $D=2+4=6$. The Planck mass
\begin{equation}\label{MPl}
M^2_{Pl}=m_{D}^{D-2}\int d^2 y\sqrt{|G(y)|}f'(R_2(y))
\end{equation}
depends on the specific features of the stationary metric $G_{ab}(y)$. Notice that the parameters $a_k$ can be fine-tuned to make the cosmological $\Lambda$-term small enough.

Varying the metric $G_{ab}(y)$ and neglecting terms proportional to the small parameter $\epsilon(x,y)$, we get
\begin{equation}\label{eq2}
f'(R)R_{ab}-\frac{1}{2}f(R)G_{ab}-\nabla_{a}\nabla_{b}f'+G_{ab}\square f'=0.
\end{equation}
Taking the trace of (\ref{eq2}), we have
\begin{equation}\label{eqtrace}
\frac{1}{\sqrt{G}}\partial_a\sqrt{G}G^{ab}\partial_b\frac{df}{dR}=f-R\:\frac{df}{dR}\ ,
\end{equation}
where $a$ and $b$ are the indices of the two coordinates in the extra space.

If we assume a spherical-like metric for the extra space,
\begin{equation}\label{sphereMetrix}
ds^2 = r^2(\theta)(d\theta^2 + \sin^2\theta\:d\phi^2),
\end{equation}
the scalar curvature will be expressed via the function $r(\theta)$:
\begin{equation}\label{Rofr}
R(\theta)=\frac{2}{r^4\sin\theta}(-r'r\cos\theta+r^2\sin\theta+(r')^2\sin\theta-rr''\sin\theta),
\end{equation}
where the prime means differentiation with respect to $\theta$.

Eqs.~(\ref{eqtrace}) and (\ref{Rofr}) can be viewed as a system of second-order ordinary differential equations (ODE) for the functions $r(\theta)$ and $R(\theta)$. We solved this system numerically in the segment $0\le\theta\le\pi$,
with boundary conditions at $\theta=0$ (see below). Notice that the case of the maximal symmetric extra space corresponds to $R(0)=R_*$, $r(0)=\sqrt{2/R_*}$, $r'(0)=r''(0)=0$.
The function $f(R)$ was chosen in the form
\begin{equation}\label{fR}
f(R)=U_1(R-R_0)^2+U_2,
\end{equation}
where $U_1$, $U_2$, and $R_0$ are constants.

\begin{figure}
\begin{center}
	\includegraphics[scale=0.5]{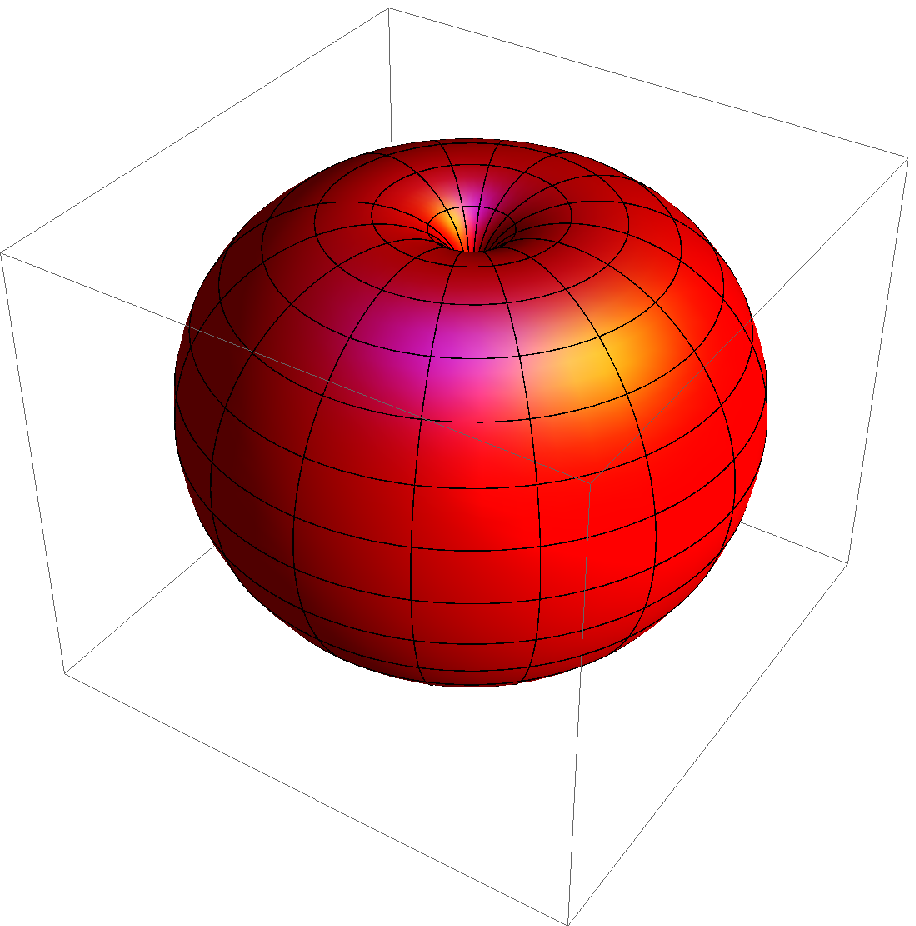}
    \hspace{10mm}
	\includegraphics[scale=0.5]{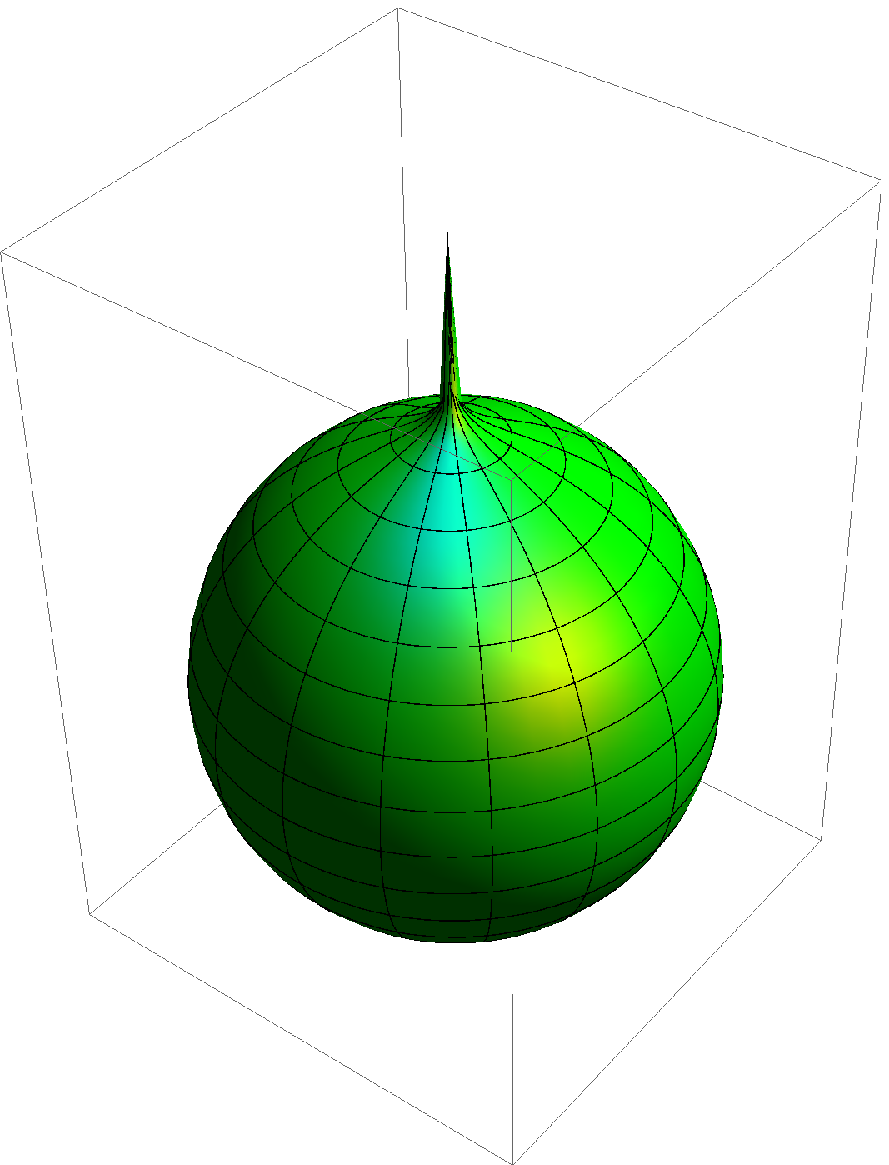}
\end{center}
	\caption{The extra space configurations of the ``apple'' type at $r''(0)=0.03$ (left panel) and of the ``onion'' type at $r''(0)=-0.01$ (right panel).}
\label{fig:fig1}
\end{figure}

Some typical results of our numerical calculations are shown in Fig.~\ref{fig:fig1}.
For $U_2=0$, depending on the value of $r''(0)$, we get configurations of the ``apple'' type (at $r''(0)>0$) and of the ``onion'' type (at $r''(0)<0$). Note that the ``apple'' type geometries have also been discussed in \cite{gogberashvili}. Fig.~\ref{fig:fig1} demonstrates that the solutions develop a singularity when $\theta\to\pi$.

Due to the symmetry of the problem, we can impose boundary conditions at $\theta=\pi$. Then we obtain a singularity at $\theta\to 0$. We can investigate the asymptotic behavior of the functions $r(\theta)$ and $R(\theta)$ at $\theta\to 0$. To this end we assume
\begin{equation}
r(\theta)=\frac{C}{\theta^b}, \quad b>0,\quad\theta \rightarrow 0.
\end{equation}
Inserting it in the system (\ref{eqtrace}), (\ref{Rofr}) we get:
\begin{equation}
r(\theta)=\frac{C}{\sqrt{\theta}}, \quad R(\theta )=\frac{5}{3C^2}\:\theta , \quad \theta \rightarrow 0,
\end{equation}
where $C=\left(\frac{3}{10}\left(\frac{U_2}{U_1}+R_0^2\right)\right)^{-1/4}$.

\section{Domains with a deformed extra space}
\label{S:3}

Consider a domain {\cal U} of the main space with the deformed extra space of the apple type or the onion type. Suppose that outside the domain {\cal U}
each point of the main space has the extra dimensions compactified into
the ideal spheres. Mass of such a domain for a distant observer can be estimated as
\begin{equation}
M\simeq (\rho-\rho_0)L^3,
\end{equation}
where
\begin{equation}
\rho=m_D^4\pi\int r^2(\theta)\sin\theta f(R(\theta))d\theta,\quad
\rho_0=m_D^4\pi\int r_*^2\sin\theta f(R_*)d\theta
\end{equation}
are the vacuum energy densities inside and outside the domain {\cal U}, respectively; $L$ is the typical size of {\cal U}.

It is natural to suppose that the minimal size of the domain {\cal U} is of order of the extra space size, i.e.\ less than $10^{-18}$ cm. In this case the mass of such a particle-like object can vary from zero to a few TeV or more. Notice that effects related to geometry of the extra dimensions can be interpreted as weakly interacting particles of the dark matter \cite{panico01,kahil01}.

Finally, we estimate the interaction cross section of such a particle-like object with the ordinary matter within the non-relativistic quantum mechanics \cite{LL}. Suppose that the domain size is of order of the extra space size, $L\sim l\sim 10^{-18}$ cm. We find the scattering cross section of the particle of mass $m$ in the Born approximation on the potential
\begin{equation}
V(x)=V_0\exp{(-x^2 /l^2)}, \quad V_0= l^{-1}.
\end{equation}
Then the scattering cross section:
\begin{equation}
\sigma \simeq \pi^2 m^2 V_0^2 l^6 \sim \pi^2 m^2 l^4 \lesssim 10^{-43}\ \mbox{cm}^2.
\end{equation}
This estimate agrees with the existing observational constraints.

\section{Conclusion}
\label{S:4}

We have considered stationary geometries of two-dimensional manifolds with the sphere-type topology. The stability of geometry in the absence of matter is a consequence of the non-linearity of the initial Lagrangian. As an example we discussed the Lagrangian being a second-power polynomial in the scalar curvature. The numerical solution of the corresponding Cauchy problem showed that there is a class of sphere-type metrics. This class consists of two sub-classes with different behavior at the singular point $\theta=\pi$. The first subclass --- the configurations of the ``onion'' type --- has a rather simple asymptotic behavior at the singular point. This asymptotic was investigated analytically. The second subclass --- the configurations of the ``apple'' type --- has a more complex structure and strongly depends on the initial conditions.

We have also shown that domains of our four-dimensional space-time with the deformed extra space at each point can be interpreted as particles of the dark matter. Such particles interact with the ordinary matter only gravitationally, their mass can vary depending on the parameters of the model and the initial conditions. The estimate of the interaction cross section of these dark matter particles with the nucleons gives a value of the order of $10^{-43}$ cm$^2$, that does not contradict the existing constraints.

In this work we have estimated the mass of the dark matter particle. To obtain its precise value it is necessary to find a stationary solution of the Einstein equation in 4+2 dimensions. We are going to study this question in the future.

\section*{Acknowledgements}
\label{S:5}

This work was supported by the Ministry of Education and Science of the Russian Federation, Project No.~3.472.2014/K. The work of V.~A.~Gani was also supported by the Russian Federation Government under Grant No.~NSh-3830.2014.2. The authors are grateful to K.~A.~Bronnikov and V.~I.~Dokuchaev for useful discussions, and also to A.~I.~Bolozdynya and M.~D.~Skorokhvatov for their interest to this work.


\begin{thebibliography}{}


\bibitem[Cianfrani (2008)]{2008IJMPD..17..785C}
Cianfrani~F. and Montani~G. Low-energy sector of eight-dimensional general relativity: electroweak model and neutrino mass. {\it Int.~J.~Mod.~Phys.~D.} 2008;17:785.

\bibitem[Mazumdar (2011)]{2011PhR...497...85M}
Mazumdar~A. and Rocher~J. Particle physics models of inflation and curvaton scenarios. {\it Phys.~Rep.} 2011;497:85-215.

\bibitem[Bolokhov (2011)]{2011PhRvD..84d4015B}
Bolokhov~S.~V., Bronnikov~K.~A., Rubin~S.~G. Extra dimensions as a source of the electroweak model. {\it Phys.~Rev.~D.} 2011;84:044015.

\bibitem[Linde (1991)]{1991PhLB..259...38L}
Linde~A. Axions in inflationary cosmology. {\it Phys.~Lett.~B.} 1991;259:38-47.

\bibitem[Linde (1990)]{Lindebook}
Linde~A.~D. {\it Particle Physics and Inflationary Cosmology.} Switzerland: Harwood Academic Publishers. 1990.

\bibitem[Bamba (2014)]{nojiri01}
Bamba~K., Makarenko~A.~N., Myagky~A.~N. {\it et al.} Bounce cosmology from F(R) gravity and F(R) bigravity. {\it J.~Cosmol.~Astropart.~Phys.} 2014;01:008.

\bibitem[Soko{\l}owski (2007)]{sokolowski01}
Soko{\l}owski~L.~M. Metric gravity theories and cosmology: II. Stability of a ground state in f(R) theories. {\it Class.~Quant.~Grav.} 2007;24:3713.

\bibitem[Gogberashvili (2007)]{gogberashvili}
Gogberashvili~M., Midodashvili~P., Singleton~D. Fermion generations from ``apple-shaped'' extra dimensions. {\it JHEP.} 2007;08:033.

\bibitem[Panico (2008)]{panico01}
Panico~G., Pont{\'o}n~E., Santiago~J. {\it et al.} Dark matter and electroweak symmetry breaking in models with warped extra dimensions. {\it Phys.~Rev.~D.} 2008;77:115012.

\bibitem[Kahil (2009)]{kahil01}
Kahil~M.~E and Harko~T. Is dark matter an extra-dimensional effect? {\it Mod.~Phys.~Lett.~A.} 2009;24:667.

\bibitem[Landau (1977)]{LL}
Landau~L.~D. and Lifshitz~E.~M. {Quantum Mechanics. Non-relativistic theory. Volume 3 of Course of Theoretical Physics}. Pergamon Press. 1977.

\end{thebibliography}
\end{document}